\documentclass[11pt,twoside]{article}  
\usepackage{asp2006}
\usepackage{adassconf}

\begin{document}   

%
%

\paperID{P3.11}

%

\title{Middleware for Data Visualization in VO-enabled Data Archives}

%
%
%
%
%

\markboth{Zolotukhin \& Chilingarian}{Middleware for Visualization in Data 
Archives}

%
%
%
%

\author{Ivan Zolotukhin \altaffilmark{1}}
\author{Igor Chilingarian \altaffilmark{1,2}}

\altaffiltext{1}{Sternberg Astronomical Institute, Moscow State University, 13 Universitetsky prospect, Moscow, 119992, Russia}
\altaffiltext{2}{Observatoire de Paris-Meudon, VO-Paris Data Centre; LERMA, UMR~8112, 61 Av. de l'Observatoire, Paris, 75014, France}


%

\contact{Ivan Zolotukhin}
\email{iz@sai.msu.ru}

%
%
%

\paindex{Zolotukhin, I.}
\aindex{Chilingarian, I.}     



\begin{abstract}          
We present a middleware for visualization and exploration of complex
datasets in a VO framework, that performs interaction between data archives
and existing VO client applications using PLASTIC. It comprises: (1)
PLASTIC-enabled Java control applet, integrated into archive web-pages and
interacting with VO applications; (2) cross-browser compatible JavaScript
part managing PLASTIC-aware VO Clients (launch, data manipulation) by means
of Java LiveConnect. This (or similar) solution is an essential for the new
generation VO-enabled data archives providing access to complex
observational and theoretical datasets (3D-spectroscopy, N-body simulations,
etc.) through web-interface. Thanks to PLASTIC capabilities it is possible
to start all necessary client software with a single-click in the archive
query result page in a web-browser. This simplifies the scientific usage of
the VO resources and makes it easy even for users with no experience in the
VO technologies.
\end{abstract}


\section{Introduction}
After several years of intensive development, the International Virtual
Observatory has accumulated numerous useful standards, tools, and access
protocols allowing users to deal with the astronomical data of almost any
kind. The fact that many software products within the Virtual Observatory
follow IVOA conventions and standards opens a very promising perspective
to attempts of integrating heterogeneous VO tools within end-user's unique
environment. One can glue VO tools developed by third parties, providing
various feature sets into a VO system based on intercommunication between
them, that will dramatically improve usability and/or feature list
available to scientific users.

Being somewhat in a contradiction to the ``classical'' development schema, this
approach brings all the power of the VO initiatives to data centres and
other service providers involved in development of new services for
astronomical needs.

The idea described here is a cornerstone of the approach to visualization of
complex observational and theoretical datasets. We have implemented it in
several VO-enabled data archives significantly increasing the value that
``usual'' archive can bring to an astronomer.

\section{Middleware Design}
The typical design of a \textit{classical} data archive implies the 
end-user to:
\begin{enumerate}
\item submit a query on a web page to search for a data needed
\item locate a dataset in the search results
\item download or put it to a VO storage service (e.g. VOStore)
\item start a (VO) tool of his choice and open the file inside it for
scientific analysis
\end{enumerate}

We propose another way of interaction between a user and a data archive by
putting a middle layer between VO-enabled tools residing on a researcher's
PC, and a data archive itself. We use the full advantage of PLatform for 
Astronomical Tools InterConnection (PLASTIC), a prototype of an application
messaging protocol, based on XML-RPC.

Our middleware eliminates some unnecessary steps between the data
discovery and analysis phases, significantly simplifying the usage of
VO-enabled data archives by end-users. It comprises:
\begin{enumerate}
\item PLASTIC-enabled Java control applet, integrated into archive web pages
and interacting with VO applications (e.g. CDS Aladin, ESA VOSpec, TOPCAT); 
\item Cross-browser compatible JavaScript part managing PLASTIC-aware VO
Clients (launch, data manipulation, etc.) by means of Java LiveConnect.
\end{enumerate}

The principal design of the entire middle layer is the following. Java
applet \textit{App\_control\_applet} is embedded into the web pages and
started in background once the user looks through the archive search
results. The applet waits for a PLASTIC hub to be started and connects to it
as soon as it has become available. Since then \textit{App\_control\_applet}
is able to check if a particular VO client application is connected to the
PLASTIC hub and control its behaviour by means of PLASTIC messages; for
instance an applet can send the data (or instructions to download them) to a
proper application.

\section{Component Details}

\textit{App\_control\_applet} has an interface to JavaScript (via Java
LiveConnect) enabling a web-browser to:
\begin{enumerate}
\item recognize if all the necessary VO tools have been launched
\item launch the desired VO tool by means of Java WebStart
\item send the data requested by the user through archive query interface 
directly to a dedicated VO application
\item control some other aspects of the VO tools behaviour which are manageable
via PLASTIC
\end{enumerate}

This creates a ``bridge'' between server and client sides, and the server
side is able control processes taking place at the client, simplifying user
communication with the archive. From the user's point of view it means there
is no need anymore to launch the VO clients and download files separately:
all these actions can be completed by a single mouse-click from any modern
web browser.

\begin{figure}[t]
\epsscale{0.90}
\plotone{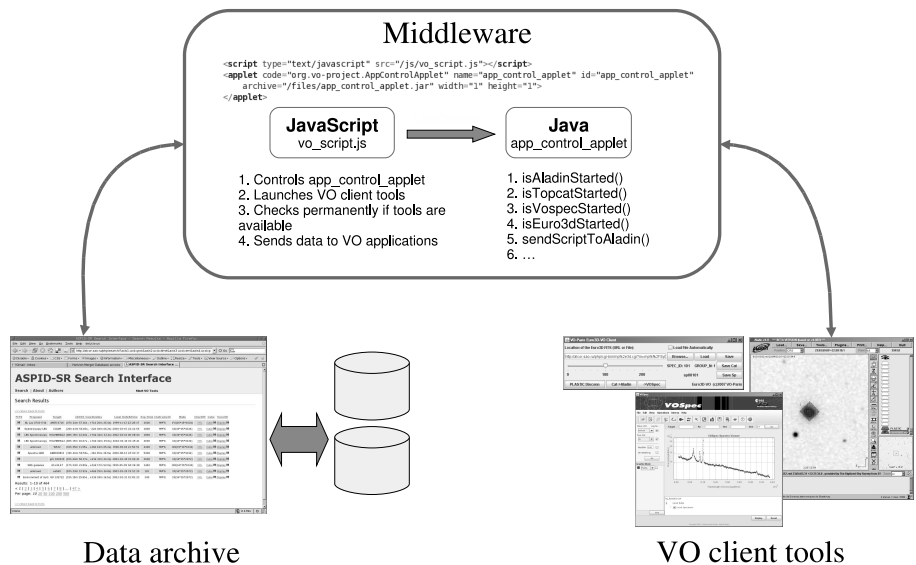}
\caption{Structure of the middleware.}
\label{P3.11_1}
\end{figure}

The key advantages of the middle layer are the following:
\begin{enumerate}
\item It is possible to start all necessary client software with a
single-click in the archive query result page in a web-browser. 
\item It does not require severe changes in the data archive implementation: 
only two lines of HTML code are required to embed the middleware in the pages
containing query results.
\item It works in any modern web browser.
\end{enumerate}

The prototypes of the data archives with the integrated middleware can be
accessed at the following web sites:
\begin{itemize}
\item ASPID-SR (Chilingarian et al. 2007), providing access to the
fully-reduced 3D spectral datasets, obtained with the Russian 6-m telescope
(see technical details in Zolotukhin et al. 2007)
\item The Horizon GalMer database, containing results
of N-Body simulations of galaxy mergers (Di Matteo et al. 2007a,b)
\end{itemize}

Details on these implementations and descriptions of interaction between
particular VO tools are given in Chilingarian \& Zolotukhin (this volume).

\section{Summary}
The presented solution for simplifying retrieval of and manipulation with
the datasets available through VO-enabled data archives is proved to be
versatile by two implementations providing access to data of completely
different origin. This (or similar) solution is an essential for modern new
generation data archives. It greatly simplifies the scientific usage of the
VO resources and makes it easy even for astronomers with no experience in
the VO technologies.

\acknowledgments
Authors wish to thank ADASS organizing committee for the financial support
provided. Travel of IZ is also supported via RFBR grant 07-02-08846.


\end{document}